# Does a 'glycine sodium nitrite' crystal exist?

Kiran T. Dhavskar and Bikshandarkoil R. Srinivasan

Department of Chemistry, Goa University, Goa 403206, India

Email: srini@unigoa.ac.in  Telephone: 0091-(0)832-6519316; Fax: 0091-(0)832-2451184

**Abstract**

The glycine sodium nitrite crystal reported by Khandpekar and Pati in the paper entitled, 'Synthesis and characterisation of glycine sodium nitrite crystals having non linear optical behaviour' Opt. Commun. 285 (2012) 288-293 is actually γ-glycine. In addition, we show that glycine barium-ammonium nitrate (GABN), glycine sodium–zinc sulfate (GSZS), glycine barium calcium nitrate, glycine acetamide and glycine dimer are dubious crystals.

**Introduction**

The simplest amino acid glycine exhibits polymorphism and three different modifications namely α- or β- or γ-glycine are well documented with the β- or γ- forms crystallizing in non-centrosymmetric space groups [1-3]. The α- and γ-forms are commercially available, while the β-form is metastable. In accordance with its achiral nature, a majority of the known structurally characterized compounds of glycine are centrosymmetric [5-9]. In spite of this, glycine has been chosen by many research groups as a precursor material for new nonlinear material synthesis. The inappropriate choice of glycine as a precursor for NLO crystal work can be evidenced by the several improperly characterized compounds of glycine in the literature [10-22], many of which were correctly formulated later (Table 1). In many cases the so called novel NLO crystals based on glycine were proved to be either α- and γ-forms of glycine. A scrutiny of the improperly characterized NLO crystals listed in Table 1 reveals that all such compounds do not have a proper chemical formula and are referred to by unusual names and abbreviated by strange codes.

Table 1 List of improperly characterized NLO crystals based on glycine

| No | NLO crystal initially claimed | Actual crystal | Ref |
|---|---|---|---|
| 1 | 1,3 Diglycinyl thiourea | α-Glycine | 10 |
| 2 | Triglycine acetate (TGAc) | α-Glycine | 10 |
| 3 | Diglycine nitrate (DGN) | α-Glycine | 10 |
| 4 | Bis-glycine sodium nitrate (BGSN) | α-Glycine | 13 |
| 5 | Bisglycine hydrogenchloride (BGHC) | γ-Glycine | 10 |
| 6 | $C_2H_{11}NO_9KCl$ | γ-Glycine | 10 |
| 7 | Glycine lithium chloride (GLC) | γ-Glycine | 11, 12 |
| 8 | Glycine barium nitrate potassium nitrate (GBNPN) | γ-Glycine | 21 |
| 9 | Glycine hydrobromide | Diglycine hydrobromide | 10 |
| 10 | Bis-glycine maleate (BGM) | Glycinium hydrogen maleate | 19 |
| 11 | Glycine picrate (GP)  $GlyH·C_6H_3N_3O_7$ | Glycine glycinium picrate $Gly·GlyH·C_6H_3N_3O_7$ | 14-16 |
| 12 | N-acetyl glycine phosphite (AGPI) | Glycinium phosphate | 20 |
| 13 | Glycine barium dichloride  $GlyBaCl_2$ | $Gly_2·BaCl_2·H_2O$ | 10 |
| 14 | Tetra glycine barium chloride (TGBC) | $Gly_2·BaCl_2·H_2O$ | 10 |
| 15 | Glycine zinc sulfate (GZS)  $Gly·ZnSO_4·7H_2O$ | $Gly·ZnSO_4·5H_2O$ | 10 |
| 16 | Glycine zinc chloride (GZC) | Diglycine $ZnCl_2$ dihydrate $Gly_2·ZnCl_2·2H_2O$ | 10 |
| 17 | Glycine ammonium oxalate (GAO) | $(NH_4)_2C_2O_4·H_2O$ | [22] |
| 18 | $2Gly·HF·HCl$ glycine–chloride–fluoride (GCF) | $2Gly·HCl$ | 17 |
| 19 | $2Gly·H_2SO_4·HNO_3$ α-glycine sulpho-nitrate | $Gly·HNO_3$ | 17 |
| 20 | $3Gly·KIO_3$  glycine potassium iodate | Unspecified* | 17 |
| 21 | $3Gly·KNO_3$  glycine potassium nitrate (GPN) | Unspecified* | 17 |
| 22 | Glycine hydrogen potassium fluoride (GHKF) | γ-Glycine | 18 |
| 23 | Glycine sodium–barium nitrate $6Gly·NaNO_3·Ba(NO_3)_2$ (GSB) | γ-Glycine | 18 |
| 24 | Glycine sodium potassium nitrate $Gly·NaNO_3·KNO_3$(GSPN-1) | γ-Glycine | 18 |
| 25 | $6Gly·NaNO_3·KNO_3·2HCl$ (GSPN-2) | γ-Glycine | 18 |
| 26 | Glycine sodium nitrite (GSNi) **1** | γ-Glycine | This work |
| 27 | Glycine ammonium-barium nitrate (GABN) **2** | γ-Glycine | This work |
| 28 | Glycine sodium–zinc sulfate (GSZS) **3** | α-Glycine | This work |
| 29 | Glycine barium calcium nitrate (GBC) **4** | Barium nitrate | This work |
| 30 | Glycine acetamide (GA) **5** | α-Glycine | This work |
| 31 | Glycine dimer **6** | α-Glycine | This work |

Abbreviations used: Gly = glycine; *Reported data for the NLO crystal not in agreement for proposed composition and structure. For details see[17].

During a literature survey of NLO crystals based on glycine, we came across compounds of glycine containing one or more *s*-block metal in the same crystal for example glycine sodium nitrite (GSNi) [23], glycine barium-ammonium nitrate (GABN) [24], glycine sodium–zinc sulfate (GSZS) [25], glycine barium calcium nitrate (GBC) [26]. In contrast to the well-documented oxidising behaviour of nitrite towards the primary amine group, the claim of isolating crystals of glycine sodium nitrite appeared not only unusual but the reported elemental analytical data were incompatible with the proposed chemical formula. In view of the importance of the *s*-block metal based amino acid compounds, we have reinvestigated all such compounds of glycine [23-28] to identify the correct formulae of such NLO crystals. The results are described herein.

**Results and discussion**

*Synthetic aspects of the crystal growth studies*:

The recently reported crystal growth of glycine sodium nitrite (GSNi) **1**, glycine ammonium-barium nitrate **2**, glycine sodium–zinc sulfate (GSZS) **3**, glycine barium-calcium nitrate (GBC) **4**, glycine acetamide **5**, glycine dimer **6**, [23-28] is reinvestigated in order to unambiguously characterize the crystalline product. The reinvestigation was undertaken due to the fact that none of these so called NLO crystals were characterized by any rigorous single crystal structure refinement method, but only based on X-ray powder diffraction (compounds **1**-**4**) and by unit cell data in the case of glycine acetamide with an unusually small beta angle and an inappropriate space group for glycine dimer. In all these cases it appeared that the formulation of these so called novel NLO crystals was based on an assumption that mixing up of a few reagents in water will result in crystallization of a desired product crystal. Such assumptions can be evidenced in the papers of Khandpekar et al [23-26] reporting crystal growth by writing some unusual chemical reactions disregarding the chemistry of the reagents employed for the crystal growth, for example the growth of a so called glycine sodium nitrite by reaction of glycine with sodium nitrite in a 3:1 mole ratio.

*Glycine sodium nitrite is pure γ-glycine*

While it is not clear as to why a crystal obtained by using three moles of glycine is called glycine sodium nitrite, we found a claim of isolation of a crystalline solid containing both an amino acid and nitrite just by mixing two reactants in water, to contradict the known chemistry of nitrite group towards amino acids. In order to verify the reported claim we reinvestigated the reaction by performing it under the reported conditions namely reaction of three moles of glycine with sodium nitrite in water and analyzed the isolated crystals (compound **1**) for the presence of Na and nitrite by standard chemical tests [29] and also recorded its infrared (IR) spectrum. The tests revealed the absence of both sodium and nitrite. More interestingly the IR spectrum of **1** was identical to that of pure γ-glycine [Fig. 1] showing that it is γ-glycine and not any so called glycine sodium nitrite. The formation of γ-glycine is not at all surprising because the reaction between α-amino acid and aqueous nitrite in which the primary amine group is oxidized to $N_2$ resulting in the formation of the corresponding α-hydroxy acid as the product (Scheme 1) is well documented in all standard biochemistry text books [30].

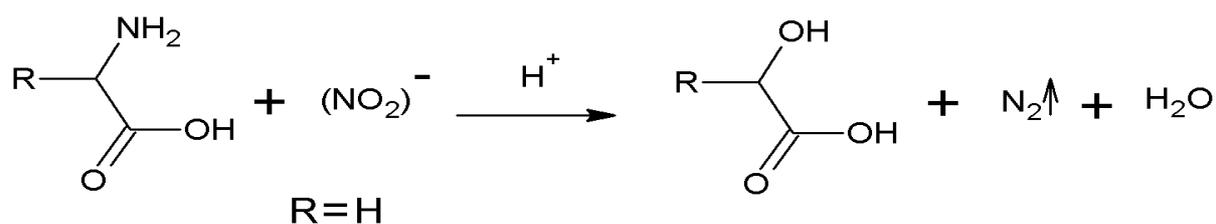

Scheme 1

Addition of sodium nitrite into a solution of glycine (pH of 0.2 M solution = 4.0 [31]) results in the formation of nitrous acid which acts as a strong oxidizing agent accepting electrons from the amino group of the amino acid (reducing agent) with the evolution of nitrogen. As three moles of glycine (excess) were used per mole of sodium nitrite, a major portion of the unreacted excess glycine was obtained as the product as evidenced by our yield. In their crystal growth of glycine sodium nitrite, the authors had reported brisk effervescence but failed to take this into consideration for the evolution of $N_2$. More interestingly, the formation of γ-glycine as the only product is further substantiated from the reported elemental analytical data of Khandpekar et al (C, 32.14; H, 6.94; N, 18.70; O, 41.78) which are in good agreement with the calculated values for glycine (Table 2). In this context, we regret to point out that the theoretically calculated % for the proposed formula (NH$_2$CH$_2$COOH)$_3$·NaNO$_2$ of the authors are incorrect and appear to be conveniently chosen numbers closer to the experimentally observed values.

*Other improperly characterized glycine compounds in the literature*:

A survey of Table 1 (entry Nos. 18-29) reveals that a majority of the improperly characterized glycine compounds in the literature are by the Khandpekar group, whose many earlier papers have already been severely criticised by Fleck and Petrosyan [17, 18]. It is to be noted that none of the so called NLO crystals were characterized by single crystal structure determination but mostly formulated based on an assumption that mixing up of a few reagents in water in some ratio will result in the crystallization of a desired NLO material or by incorrect interpretation of spectral data. In recent work we have shown the usefulness of infrared spectroscopy for product characterization [22], we have used this method for formulation of the so called NLO crystals glycine ammonium barium nitrate (GABN) **2**, glycine sodium–zinc sulfate (GSZS) **3**, glycine barium calcium nitrate (GBC) **4** correctly as γ-glycine, α-glycine and barium nitrate respectively which can be explained due to fractional crystallization. The infrared spectra of the products obtained are identical with the infrared spectrum of one of the starting materials (Fig. 2-5). Our correct formulation gains credence from the elemental analytical data reported by the Khandpekar group (Table 3).

In an earlier paper Krishnakumar et al [27] have claimed the growth of so called glycine acetamide crystal from an aqueous solution by the slow evaporation method and characterized it based on unit cell data as a monoclinic crystal and not by any single crystal structure determination. The dubious nature of the crystal as seen by an impossible monolclinic β angle of 11.62$^o$ called for a reinvestigation of this system. The infrared spectrum of the crystalline product from an aqueous solution containing equimolar amounts of glycine and acetamide was identical to that of an authentic α-glycine sample showing that the crystal obtained is actually α-glycine and not any glycine acetamide. The formation of pure α-glycine not only indicates that there is no chemical reaction between glycine and acetamide at ambient temperature but also explains the fractional crystallization of α-glycine with the more soluble acetamide remaining in solution. In a theoretical study Friant-Michel and Ruiz-López [32] have reported that zwitterion–zwitterion glycine dimers might be abundant in supersaturated glycine aqueous solutions a fact that has been connected with the structure of α-glycine crystals. In a recent paper Kishor Kumar et al [28] have claimed the growth of a so called novel NLO glycine dimer crystal. However this claim of growing a glycine dimer appears to be dubious as can be readily evidenced from the fact that these authors assign an unheard of 1[*P*1] space group in the orthorhombic crystal system from a powder diffraction study. A reinvestigation of the crystal growth under reported conditions indeed reveals that the crystals are pure α-glycine and not any glycine dimer.

**Experimental details**

All chemicals used namely γ–glycine (Spectrochem), sodium nitrite (Lobachemie), sodium sulphate (Thomas Baker), zinc sulphate heptahydrate (Molychem), calcium nitrate tetrahydrate (Lobachemie), barium nitrate (Molychem), ammonium nitrite (Fisher Scientific) are of reagent grade. Double distilled water was used for crystal growth. Infrared (IR) spectra were recorded in KBr matrix using a Shimadzu (IR Prestige-21) FT-IR spectrometer in the range 4000 – 400 cm$^{-1}$.

*Reinvestigation of crystal growth of a so called glycine sodium nitrite (GSNi)*

A mixture of γ–glycine (0.751 g, 10 mmol) and sodium nitrite (0.230 g, 3.3 mmol) was taken in 30 ml of hot double distilled water, stirred well to obtain a clear solution (pH=5.57). Brisk effervescence was observed. The reaction mixture was filtered and the clear filtrate was left undisturbed for crystallization. Slow evaporation of solvent maintained at room temperature resulted in the separation of transparent crystals after 5-6 days. The crystals were isolated by filtration, washed with little ice-cold water and dried in air to yield 0.421 g of crystalline product. The crystals thus obtained were labeled as **compound 1.** The details of reinvestigation of so called crystals viz. glycine ammonium barium nitrate (GABN), zinc sulphate heptahydrate (GSZS), glycine barium calcium nitrate (GBC), grown under the same reaction conditions as reported in the literature are given in supplementary material.

*Reinvestigation of glycine ammonium barium nitrate (GABN)*

A mixture of glycine (0.751 g, 10 mmol) and ammonium nitrite (0.133g, 6.6 mmol) and barium nitrate (0.435g, 6.6) was taken in 15-20 ml of double distilled water, stirred well to obtain a clear solution (pH=5.31) and 0.5% of conc HCl was added to the solution and pH was maintained at 4.4. The reaction mixture was filtered and the clear filtrate was left undisturbed for crystallization. Slow evaporation of solvent maintained at room temperature resulted in the separation of transparent crystals after 2-3 weeks. The crystals were isolated by filtration, washed with little ice-cold water and dried in air to yield 0.561 g of crystalline product. The crystals thus obtained (**compound 2**) were investigated by IR spectra. This indicates that **2** is gamma gycine with some impurity of barium nitrate. On proper washing of the crystals obtained with cold water, the more soluble glycine seperates out and we obtain the IR spectrum (Fig. 2) that resembles the barium nitrate.

*Reinvestigation of Glycine sodium–zinc sulphate (GSZS)*

A mixture of γ-glycine (0.751 g, 10 mmol), sodium sulphate (0.237 g, 6.6 mmol) and zinc sulphate heptahydrate was taken in 30 ml of double distilled water, stirred well to obtain a clear solution (pH=4.27). The reaction mixture was filtered and the clear filtrate was left undisturbed for crystallization. Slow evaporation of solvent maintained at room temperature resulted in the separation of transparent crystals after 3-4 days. The crystals were isolated by filtration, washed with little ice-cold water and dried in air to yield 0.432 g of crystalline product (**3**). This reaction yielded glycine but in a different polymorphic form namely α-glycine. This is easily understood as glycine is least soluble among three (24.99 g / 100 ml), hence crystallises out first, leaving behind the more soluble reactants zinc sulphate heptahydrate (57.7 g / 100 ml) and sodium sulphate (44 g / 100 ml) into the solution. Formation of α-glycine is confirmed based on infrared spectra (Fig. 3).

*Glycine barium calcium nitrate (GBC)*

To a solution of glycine (0.751 g, 10 mmol) in 20ml double distilled water (pH=5.58) add calcium nitrate tetrahydrate (0.394g, 6.6mmol) (pH=5.43). Then add barium nitrate (0.436g, 6.6 mmol) was taken in 30 ml of double distilled water, stirred well to obtain a clear solution. (pH=5.36) The reaction mixture was filtered and the clear filtrate was left undisturbed for crystallization. Slow evaporation of solvent maintained at room temperature resulted in the separation of transparent crystals after 3-4 days. The crystals were isolated by filtration, washed with little ice-cold water and dried in air to yield 0.253 g of crystalline product. The crystals thus obtained are labeled as **compound 4** and were analysed by IR spectra and chemical analysis. Reinvestigation of reaction 2.3 yields **compound 4** as product whose infra red spectra resembles to that of starting material barium nitrate (Fig. 4). This is expected based on fractional crystallization where solubility of barium nitrate (10.5g/100ml) is less than Glycine (24.99/100ml) and Calcium nitrate tetrahydrate (129 g / 100ml ).

*Reinvestigation of Glycine acetamide*

A mixture of γ-glycine  (0.751 g, 10 mmol) , acetamide (0.591 g, 10 mmol) was taken in 20 ml of double distilled water, stirred well for an hour  to obtain a clear solution. Slow evaporation of solvent maintained at room temperature resulted in the separation of transparent crystals after 5-6 days. These crystals (**compound 5)** upon analysis were proved to be that of pure α-glycine based on the coincidence of the spectrum with an authentic sample (Fig. 5).

*Reinvestigation of Glycine dimer*

A solution  of  γ-glycine  in  water was left undisturbed for crystallization and  the product  that was obtained on slow evaporation of solvent (**compound 6**). The Infrared spectrum of **compound 6** was identical to that of α-glycine (Fig. 5).

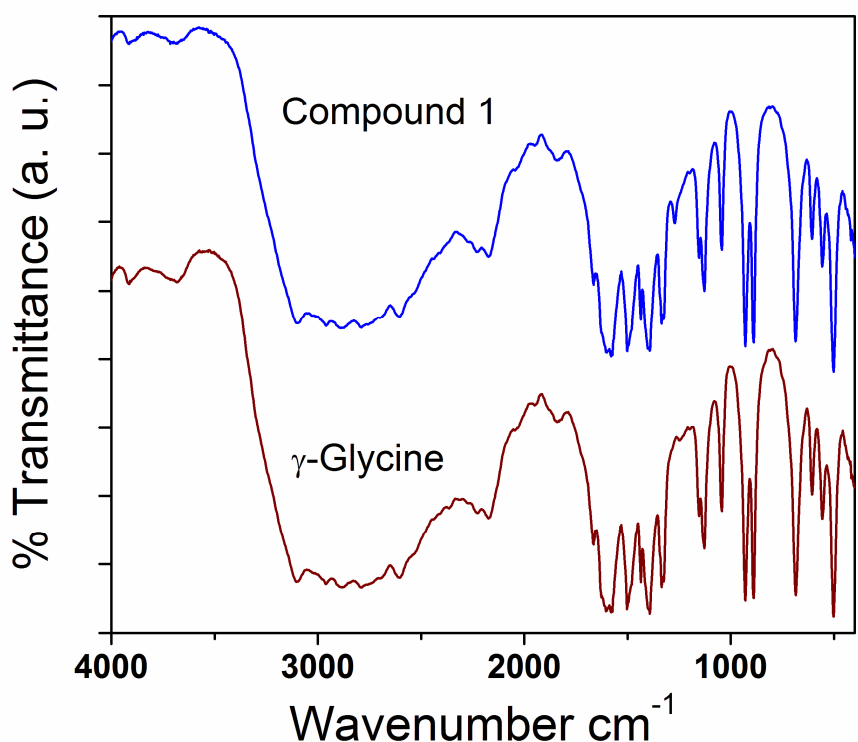

Fig.1:  The identical nature of IR spectra of compound **1** with that of γ-glycine proves a so called glycine sodium nitrite is nothing but γ-glycine.

**Table 2.** Theoretical elemental analytical data for γ-glycine and glycine sodium nitrite (GSNi) based on molecular formula

| Compound Formula | Formula weight | %C | %H | %N | %O | %Na |
|---|---|---|---|---|---|---|
| γ-glycine ($NH_2CH_2COOH$) | 75.07 | 32.00 | 6.71 | 18.66 | 42.63 | --- |
| Glycine sodium nitrite ($(NH_2CH_2COOH)_3 \cdot NaNO_2$) | 294.20 | 24.50 **(32.06)*** | 5.14 **(6.70)*** | 19.04 **(18.70)*** | 43.51 **(42.54)*** | 7.81 --- |

*__Values in bracket__ are the __wrong__ theoretical values reported by authors of [23].

**Table 3** Correct elemental analytical data calculated for the formula proposed by Khanpekar

| Compounds reported by Khanpekar et al | CHN and EDAX data of compounds | | | | | | |
|---|---|---|---|---|---|---|---|
| Glycine ammonium barium nitrate (GABN) $[Ba(NO_3)_2]_{1/2} \cdot [NH_2CH_2COOH]_3 \cdot [NH_4NO_3]_{1/2} \cdot HCl$ | %C | %H | %N | %O | %Cl | %Ba | |
| | 16.67 | 4.20 | 16.20 | 38.86 | 8.20 | 15.88 | |
| | **(31.481)*** | **(6.386)*** | **(18.237)*** | **(43.897)*** | | | |
| Glycine sodium–zinc sulfate (GSZS) $(Zn) \cdot (CH_2NH_2COOH) \cdot (Na_2) \cdot 2(SO_4) \cdot 7H_2O$ | %C | %H | %N | %O | %S | %Na | %Zn |
| | 4.76 | 3.79 | 2.78 | 53.89 | 12.71 | 9.11 | 12.96 |
| | **(15.38)*** | **(4.36)*** | **(8.91)*** | **(70.98)*** | **(0.37)*** | **(3.88)*** | **(11.63)*** |
| Glycine barium calcium nitrate (GBC) $(Ca(NO_3)_2) \cdot (NH_2CH_2COOH)_3 \cdot (Ba(NO_3)_2)_{1/2}$ | %C | %H | %N | %O | %Ca | %Ba | |
| | 13.86 | 2.91 | 16.16 | 46.16 | 7.71 | 13.21 | |
| | **(23.685)*** | **(4.922)*** | **(17.655)*** | **(53.738)*** | | | |

*__Values in bracket__ are the __wrong__ experimental values for suggested formula as reported by authors of [24-26]. The experimental values are in disagreement with the calculated data showing the dubious nature of compounds **2**-**4**. For 2 the values are in agreement for glycine.

*Glycine ammonium barium nitrate (GABN)*

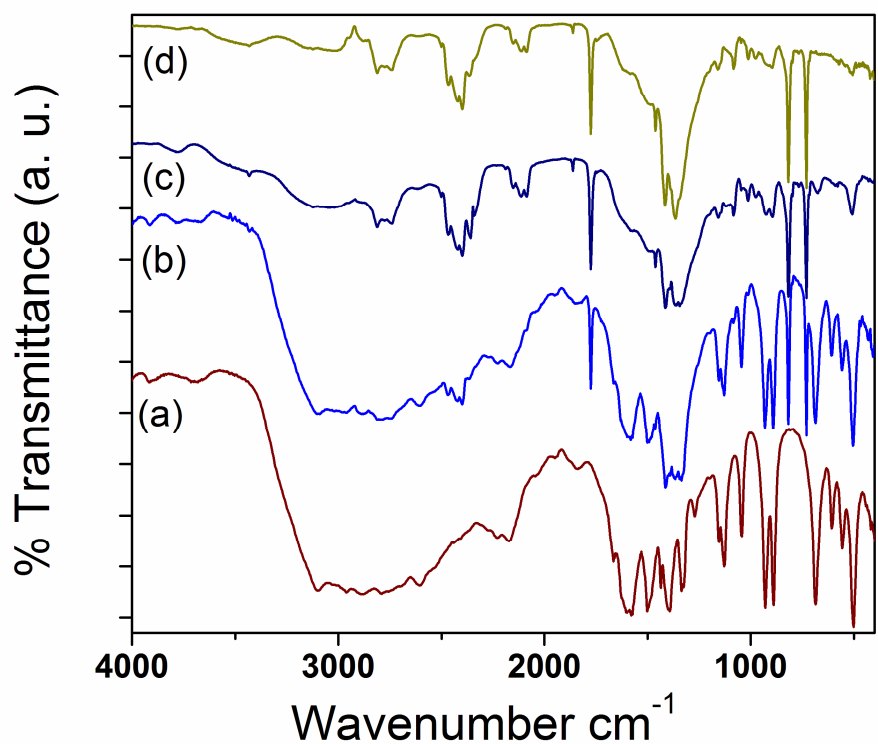

**Fig. 2** Comparative infrared spectra of (a) γ-Glycine (b) unwashed compound 4 (γ-Glycine with impurity of barium nitrate) (c) washed compound 4 (barium nitrate with trace of γ-Glycine as impurity) and (d) pure Barium nitrate

*Glycine sodium–zinc sulphate (GSZS)*

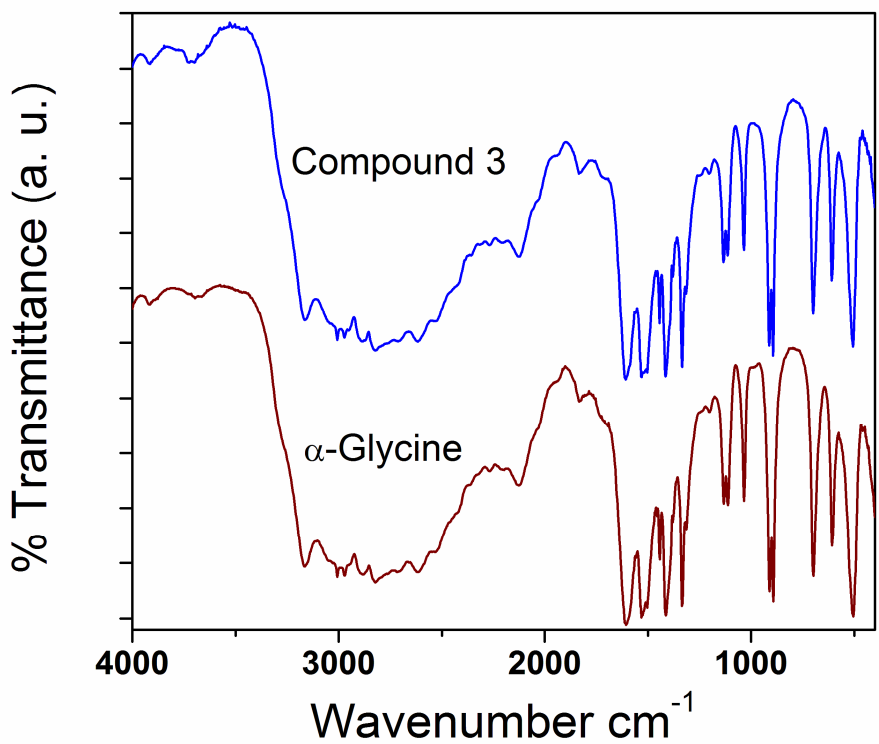

**Fig. 3** Comparative infrared spectra of **Compound 3** and pure α-Glycine

*Glycine barium calcium nitrate (GBC)*

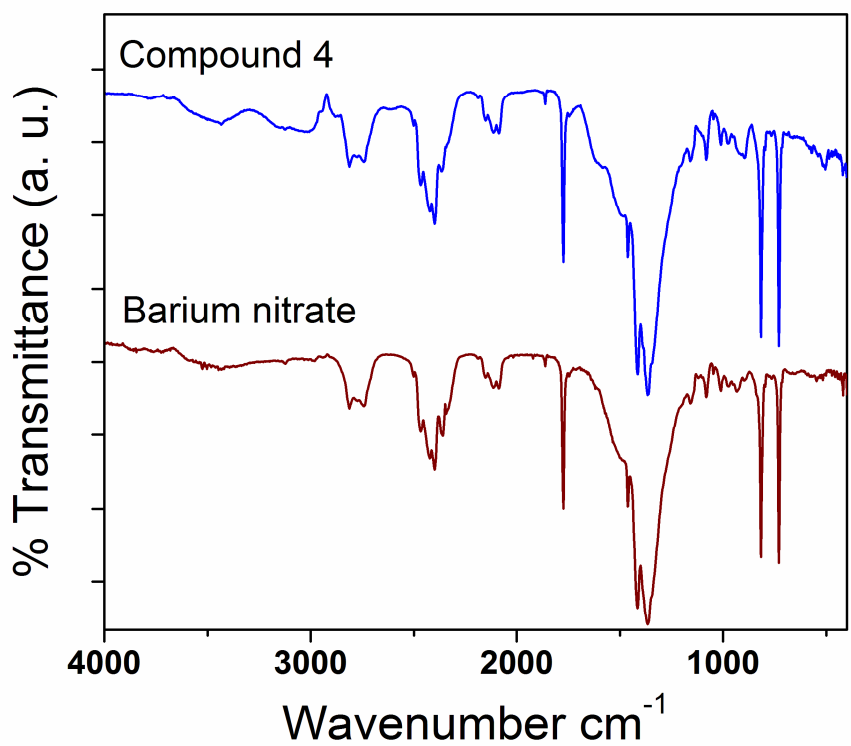

**Fig. 4** Comparative infrared spectra of **Compound 4** and pure Barium nitrate

*Glycine acetamide and Glycine dimer*

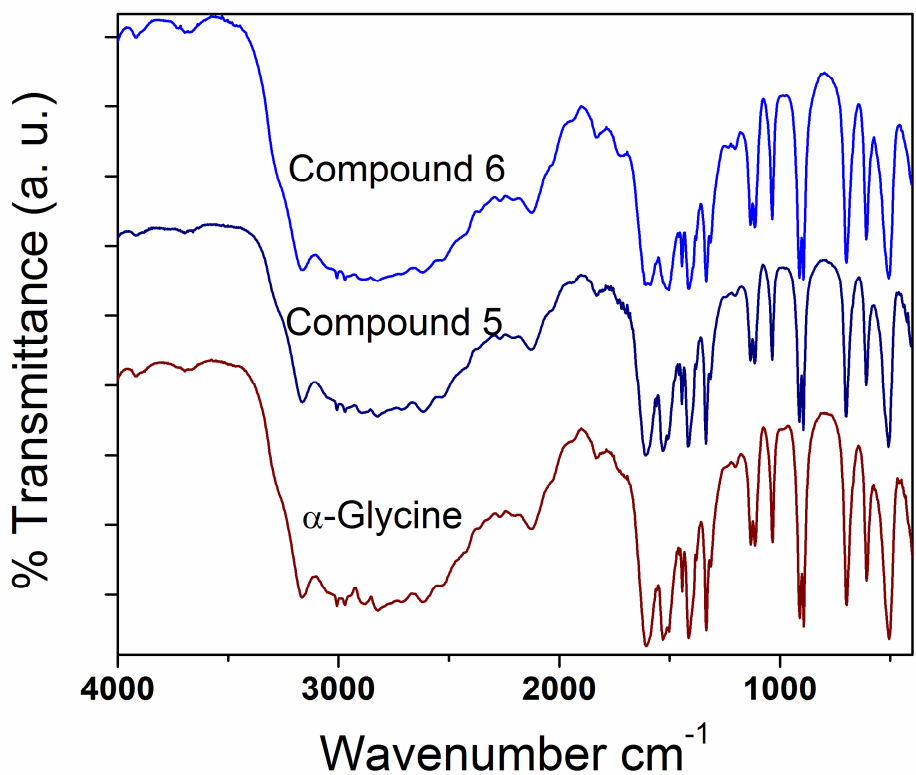

**Fig. 5** Comparative infrared spectra of **Compound 5, 6** and pure α-Glycine

**Conclusions**

In this paper we have compiled a list of several improperly characterized compounds based on glycine. Our present study reveals the risk of formulating new compounds / crystals by disregarding the chemistry of the reactants and based on an assumption that a crystal grown from a mixture of precursor materials (for example glycine and sodium nitrite) taken in a preferred ratio necessarily represents an expected compound. The main findings are as follows i) Crystal obtained from aqueous solution of sodium nitrite and glycine is pure γ-glycine ii) Infrared spectroscopy can be used to distinguish between the reactants and product in a crystal growth study. iii) All the so called novel NLO materials are actually one of the starting materials whose formation can be explained due to fractional crystallization.